\documentclass[aps,prd,reprint,showpacs]{revtex4-1}

\usepackage{acronym}
\usepackage{graphicx,psfrag}
\usepackage{amsmath,amsfonts,amssymb}
\usepackage{mathrsfs}
\usepackage{multirow,hyperref}
\usepackage{comment}
\usepackage{pifont}
\usepackage[normalem]{ulem}
\usepackage{scrextend}
\usepackage[dvipsnames]{xcolor}

\newcommand{\dd}{\mathrm{d}}

\newcommand{\ie}{\textit{i.e.}}
\newcommand{\eg}{\textit{e.g.}}
\newcommand{\pz}{\phantom{0}}
\newcommand{\SNR}{\mathrm{SNR}}

\usepackage{color}

\begin{document}

\begin{flushright}
LIGO-P1600075-v1
\end{flushright}

\title{The One-Armed Spiral Instability in Neutron Star Mergers\\ and its
Detectability in Gravitational Waves}

\author{David \surname{Radice}}
\affiliation{TAPIR, Walter Burke Institute for Theoretical Physics,
  California Institute of Technology, 1200 E
California Blvd, Pasadena, California 91125, USA}
\author{Sebastiano \surname{Bernuzzi}}
\affiliation{DiFeST, University of Parma, and INFN, I-43124
  Parma, Italy}
\affiliation{TAPIR, Walter Burke Institute for Theoretical Physics,
  California Institute of Technology, 1200 E
California Blvd, Pasadena, California 91125, USA}
\author{Christian D. \surname{Ott}}
\affiliation{TAPIR, Walter Burke Institute for Theoretical Physics,
  California Institute of Technology, 1200 E
  California Blvd, Pasadena, California 91125, USA}
\affiliation{Yukawa Institute for Theoretical Physics, Kyoto University,
  Kyoto, Japan}


\begin{abstract}
We study the development and saturation of the $m=1$ one-armed spiral
instability in remnants of binary neutron star mergers by means of
high-resolution long-term numerical relativity simulations. Our results suggest
that this instability is a generic outcome of neutron stars mergers in
astrophysically relevant configurations; including both ``stiff'' and ``soft''
nuclear equations of state. We find that, once seeded at merger, the $m=1$ mode
saturates within $\sim 10\ \mathrm{ms}$ and persists over secular timescales.
Gravitational waves emitted by the $m=1$ instability have a peak frequency
around $1-2$~kHz and, if detected, they could be used to constrain the
equation of state of neutron stars. We construct hybrid waveforms spanning the
entire Advanced LIGO band by combining our high-resolution numerical data with
state-of-the-art effective-one-body waveforms including tidal effects.  We use
the complete hybrid waveforms to study the detectability of the one-armed spiral
instability for both Advanced LIGO and the Einstein Telescope.  We conclude that
the one-armed spiral instability is not an efficient gravitational wave emitter.
Even under very optimistic assumptions, Advanced LIGO will only be able to
detect the one-armed instability up to $\sim 3\, \mathrm{Mpc}$, which
corresponds to an event rate of $10^{-7}\, \mathrm{yr}^{-1} -
10^{-4}\,\mathrm{yr}^{-1}$. Third generation detectors or better will likely be
required to observe the one-armed instability.
\end{abstract}

\pacs{
  04.25.D-,    
  04.30.Db,    
  95.30.Sf,    
  95.30.Lz,    
  97.60.Jd     
}
\maketitle

\section{Introduction}
With the recent detection of \acp{GW} from a pair of merging \acp{BH}
\cite{gwdetection}, we have entered the era of \ac{GW} astronomy.  Binary
neutron star (BNS)\acused{BNS}\acused{NS} mergers are among the targets for the
latest generation of laser-interferometer \ac{GW} detectors Advanced LIGO
\cite{advligo}, Advanced Virgo \cite{advvirgo}, and KAGRA \cite{kagra}.
The direct detection of \acp{GW} from \ac{BNS} mergers will reveal important
aspects of the physics and astrophysics of \acp{NS}. Accurate phasing
measurements of the \ac{GW} signal during the late inspiral and merger in
combination with robust theoretical predictions, \eg, \cite{hotokezaka:15,
bernuzzi:15a, hotokezaka:16}, will provide accurate and nearly model-independent
measurements of masses and radii of \acp{NS}. This will help constrain the
unknown physics of matter at supernuclear densities \cite{damour:12, read:13,
delpozzo:13, lackey:15, agathos:15, hotokezaka:16}.

Recent observations of \acp{NS} with masses $\sim 2\ M_\odot$ \cite{demorest:10,
antoniadis:13} in combination with the distribution of \ac{NS} masses in
galactic binaries, which peaks at $\sim 1.35\ M_\odot$ \cite{lattimer:12,
kiziltan:13}, suggests that the typical outcome of mergers is the formation of a
stable remnant or of a \ac{HMNS}. The latter is a metastable object that may
survive for several tens of milliseconds before collapsing to a \ac{BH}
\cite{baumgarte:00, shibata:06, baiotti:08, sekiguchi:11, bauswein:11,
hotokezaka:13, takami:14, palenzuela:15, bernuzzi:15c}.  The merger remnant is
an efficient \ac{GW} emitter \cite{bernuzzi:15c} and its gravitational radiation
has discrete features (peaks) that could be used to provide additional
constraints for the high-density part of the \ac{NS} \ac{EOS} \cite{bauswein:11,
bauswein:14, takami:14, takami:15, bernuzzi:15b, foucart:15, clark:15,
depietri:15}. However, the prospects for detecting this signal are diminished by
its high-frequency ($2-4\ \mathrm{kHz}$), which puts it outside of the
maximum-sensitivity band of current \ac{GW} detectors.

Recently, \cite{paschalidis:15, east:16} considered the merger of spinning
\acp{NS} on eccentric orbits and found that the resulting \ac{HMNS} develops an
$\ell=2,\ m=1$, one-armed spiral instability.  Because of its $m=1$ nature, this
instability results in \ac{GW} emission at half the frequency of the dominant
$m=2$ quadrupole mode, in a band of higher sensitivity for \ac{GW} detectors.
Similar instabilities have been previously identified in isolated differentially
rotating \acp{NS} models \cite{ou:06, corvino:10} and in newly-formed \acp{NS}
in core collapse, e.g., \cite{ott:05, ott:07, takiwaki:16}. There are also
strong indications of the presence of an $m=1$ instability in previous
simulations of spinning \acp{BNS} mergers in quasi-circular orbits
\cite{bernuzzi:14a, kastaun:15}. However, the impact of this instability for
\ac{GW} observation of \ac{BNS} mergers is unclear. Especially in the case of
binaries with slowly rotating or nonrotating \acp{NS} in quasi-circular orbit,
which are presumably the most common.

In this paper we present results of high-resolution \ac{NR} simulations
suggesting that the growth of an $m=1$ instability is a generic outcome of
\ac{BNS} mergers and independent of the \ac{NS} \ac{EOS}. The one-armed spiral
instability is only weakly damped and persists for several tens of milliseconds.
However, we find that the one-armed spiral instability is an inefficient emitter
of \acp{GW}. Their detection by current and near-future ground-based \ac{GW}
observatories is unlikely.

\section{Numerical Model}
We consider the \ac{GW}-driven merger of two equal mass $q=M_A/M_B=1$, $M_A =
M_B = 1.35\ M_\odot$ \acp{NS}.

We construct quasi-circular initial data in the
conformally flat approximation assuming irrotational flow \cite{gourgoulhon:01}.
We treat \ac{NS} matter as a perfect fluid and use two nuclear-theory-motivated
piecewise polytropic \ac{EOS} \cite{read:09} to close the equations of \ac{GR}
hydrodynamics.

The \ac{EOS} that we employ are designed to fit the SLy
\cite{douchin:01} and MS1b \cite{mueller:96} interaction models.  With maximum
non-rotating \ac{NS} masses of $2.06\ M_\odot$ and $2.76\ M_\odot$ respectively,
these two \ac{EOS} are representative choices of a ``soft'' and a ``stiff''
\ac{EOS}. Thermal effects during the evolution are included using a gamma-law
\ac{EOS} component  with $\Gamma_{\mathrm{th}} = 1.75$ \cite{bauswein:10}. The
initial separation between the centers of the two \acp{NS} is $50\ \mathrm{km}$,
corresponding to approximatively $9$ and $11$ orbits before merger for the MS1b
and SLy models, respectively.

The simulations are performed using the \texttt{Einstein Toolkit}
\cite{loffler:11}. For the spacetime evolution we use the Z4c formulation
\cite{bernuzzi:09} of Einstein's equations, implemented in the \texttt{CTGamma}
code \cite{pollney:09}. The \ac{GR} hydrodynamics equations are solved using the
high-order \texttt{WhiskyTHC} code \cite{radice:12, radice:13a, radice:13b}.
Our numerical grid covers a cubical region of $2048\ M_\odot \simeq 3025\
\mathrm{km}$ centered around the center of mass of the system. We enforce
reflection symmetry across the $z = 0$ plane. We use the adaptive mesh
refinement driver \texttt{Carpet} \cite{schnetter:03} to set up a grid
consisting of 7 refinement levels, with the finest ones being dynamically moved
to follow the centroids of the two \acp{NS}. The innermost refinement
level contains the \acp{NS} during the inspiral and the \ac{HMNS} after merger.
For each \ac{EOS} we perform simulations with four different resolutions having
grid spacing, in the finest level, $h = 0.25\ M_\odot, 0.2\ M_\odot, 0.15\
M_\odot$, and $0.1\ M_\odot$ (corresponding to approximatively $369\ \mathrm{m},
295\ \mathrm{m}, 222\ \mathrm{m}$, and $148$ m). For the time integration, we
use a third order strong stability-preserving Runge-Kutta method
\cite{gottlieb:01} with the Courant-Friedrichs-Lewy factor set to $0.3$.
Finally, for our analysis, we consider up to the $\ell_{\max} = 4$ multipole of
the gravitational radiation as extracted at future null-infinity $\mathcal{J}^+$
using the gauge-invariant Cauchy characteristic extraction method developed by
\cite{reisswig:09}.

\section{One-Armed Spiral Instability}
The MS1b merger results in the creation of a stable \ac{NS}, thanks to the large
maximum mass supported by this \ac{EOS}. The SLy binary forms a short lived
\ac{HMNS}. We find the survival time to be resolution-dependent with the $0.25\
M_\odot, 0.15\ M_\odot$, and $0.1\ M_\odot$ simulations showing apparent horizon
formation at approximatively $62.8\ \mathrm{ms}$, $16.3\ \mathrm{ms}$, and
$14.9\ \mathrm{ms}$ after merger, defined as the time $t_0$ when the amplitude
of the $\ell = 2,\ m = 2$ of the \ac{GW} strain peaks. We do not continue the
SLy $h=0.2\ M_\odot$ simulation until \ac{BH} formation.

In all our simulations we observe a spontaneous symmetry breaking of the system
(\eg, \cite{crawford:91}): small asymmetries due to the floating point
truncation error in our code are amplified by the turbulence generated by the
Kelvin-Helmholtz instability in the contact region between the two stars
\cite{rosswog:02, anderson:08, baiotti:08}.  This seeds physical odd-$m$
instabilities in the merger remnant.

\begin{figure*}
  \includegraphics[width=0.48\textwidth]{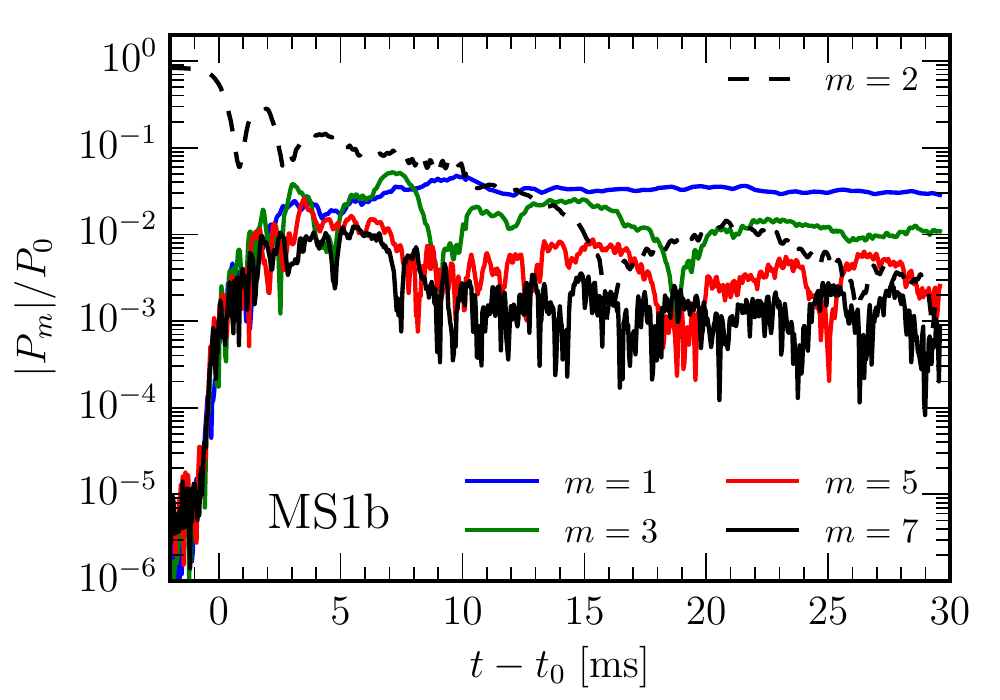}
  \includegraphics[width=0.48\textwidth]{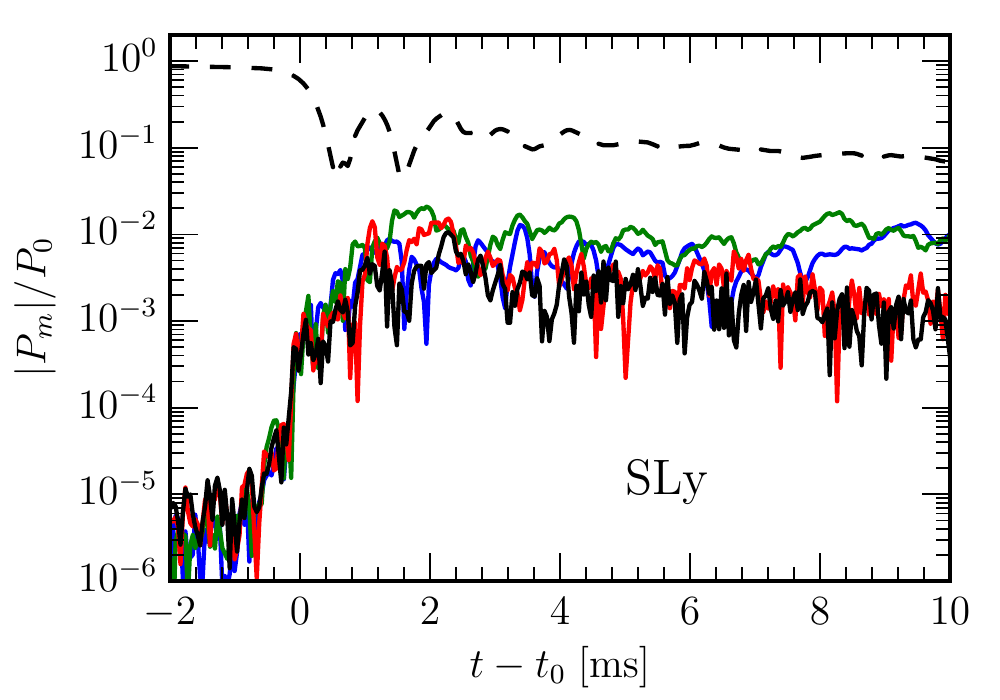}
  \caption{Normalized amplitudes of density modes on
      the equatorial plane. The symmetry-breaking odd-$m$ modes start
      to grow exponentially at the time the NSs enter into contact and
      saturate within few milliseconds.}
  \label{fig:dens_modes}
\end{figure*}

We study the development of modes violating the
$180^\circ$-rotational-symmetry ($\pi$-symmetry) of the initial data
using a modal decomposition of the density on the equatorial plane,
\begin{equation}
  P_i = \int_{\mathbb{R}^2} \rho\, W\, e^{-i m \phi}\, \sqrt{\gamma}\,
  \dd x\, \dd y\,,
\end{equation}
where $\rho$ is the rest-mass density, $\gamma$ the determinant of the
three-metric, and $W$ the Lorentz factor. We show the results of this
analysis in Figure \ref{fig:dens_modes}. Similarly to what has been
reported for spinning and/or eccentric \acp{BNS} mergers
\cite{bernuzzi:14a, kastaun:15, paschalidis:15, east:16}, we find
that, at merger, several odd-$m$ modes are seeded.  These grow
exponentially until saturation is reached, within $\sim 10\,
\mathrm{ms}$. Among these, the $m=1$ is one of the dominant modes and the
most promising for \ac{GW} detection. The $m=3$ mode is also excited
and relatively strong, especially with the SLy \ac{EOS}, but its
\ac{GW} emission is at high frequency, out of the maximum sensitivity
band of \ac{GW} detectors.

\begin{figure*}
  \includegraphics[width=\textwidth]{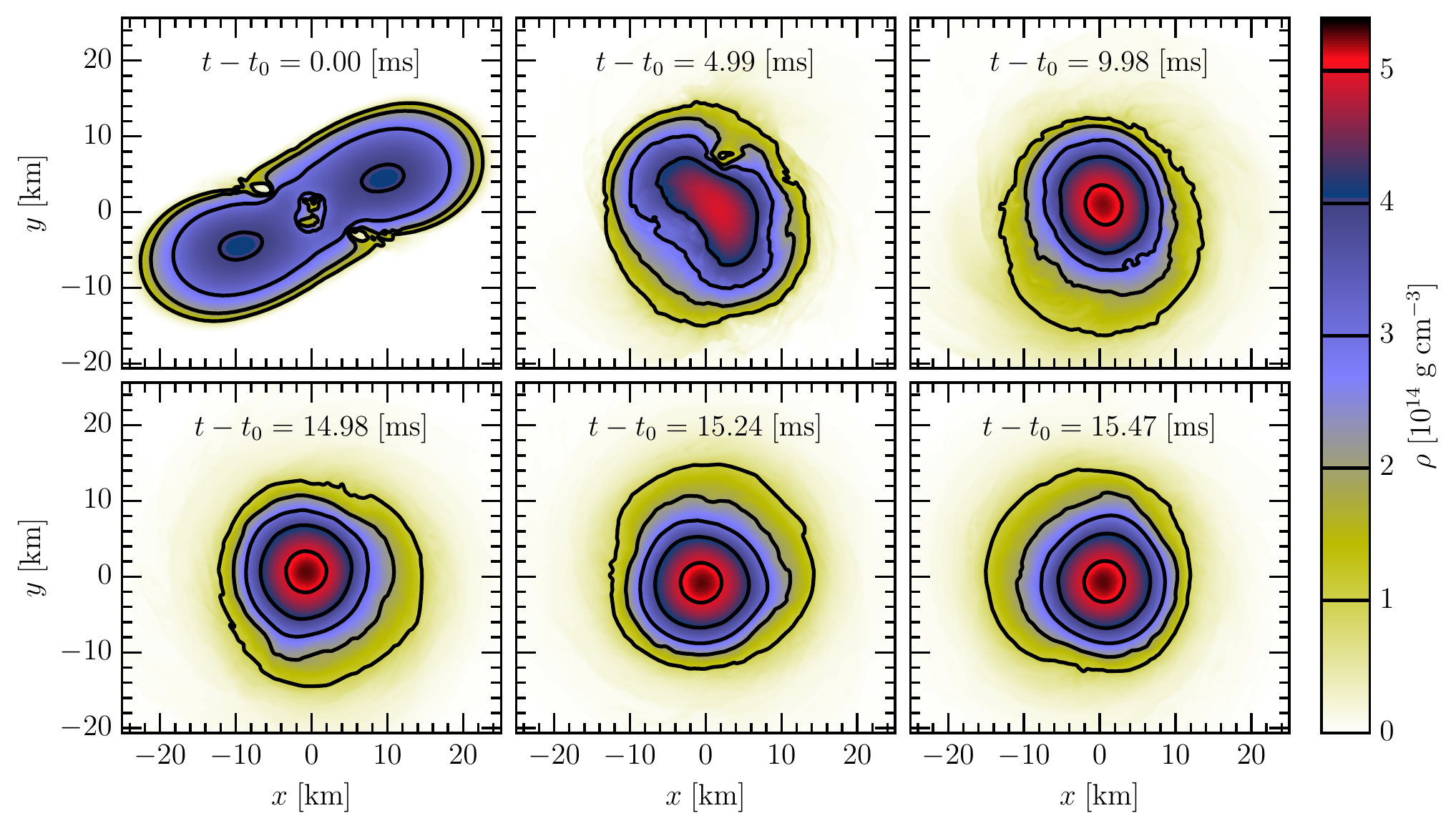}
  \caption{Color coded rest-mass density in the orbital plane for our
  highest-resolution MS1b simulation at representative times after merger
  ($t_0$).  In the upper panels: merger and development of the $m=1$
  instability.  In the bottom panels: roughly half a cycle of the saturated
  spiral mode. The $180^\circ$ rotational symmetry of the system is broken by
  hydrodynamic instabilities originating at the Kelvin-Helmholtz-unstable shear
  layer between the two stars shortly after merger.  This causes the spiral mode
  to grow.  Animations of the density in the orbital plane for both MS1b and SLy
  binaries are available as supplemental online material.}
  \label{fig:rho.xy}
\end{figure*}

Figure~\ref{fig:rho.xy} shows that for the MS1b binary with $h = 0.1\ M_\odot$,
the one-armed spiral instability appears to develop similarly to the case of
eccentric mergers \cite{paschalidis:15, east:16}.  Hydrodynamical vortices are
formed at the time of merger (under-dense regions in the first panel of
Figure~\ref{fig:rho.xy}).  These vortices subsequently migrate toward the center
where they join. This displaces the forming core of the merger remnant and
triggers the development of a spiral arm. Note that some previous studies (e.g.,
\cite{centrella:01, saijo:03}) suggested that toroidal (maximum density in a
torus around the center) rather than spheroidal (centrally condensed) stellar
structure was necessary for the one-armed instability to develop. We do not find
this to be the case: while our models exhibit slightly off-center density peaks,
they are globally spheroidal (cf.~also \cite{ott:05, ou:06, ott:07,
takiwaki:16}).

\begin{figure*}
  \includegraphics[width=0.48\textwidth]{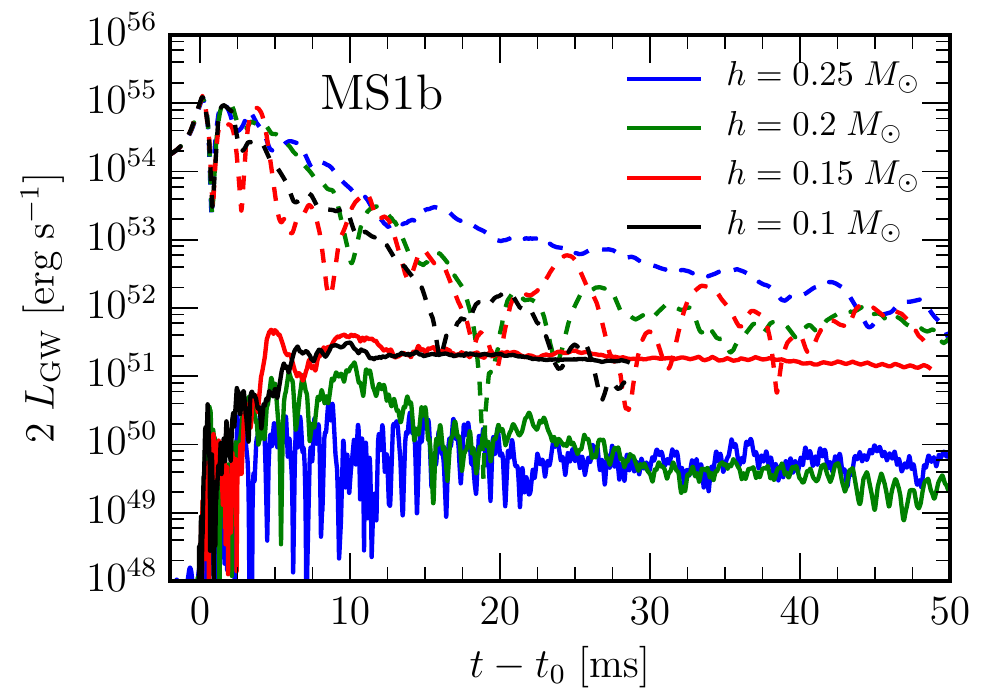}
  \includegraphics[width=0.48\textwidth]{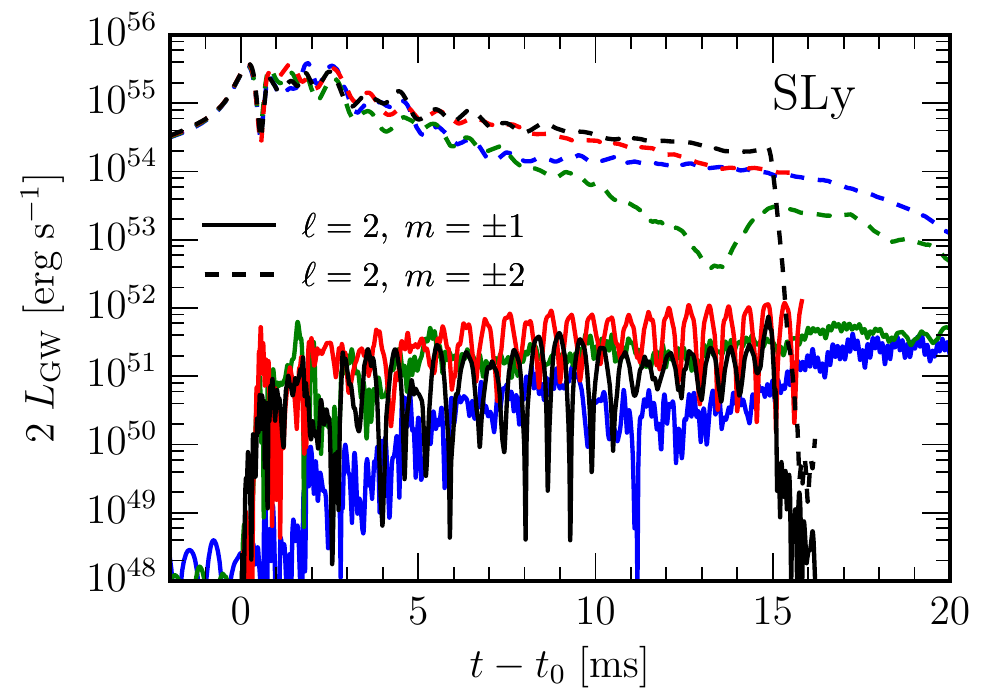}
  \caption{\ac{GW} luminosity of the $\ell = 2,\ m = 2$ and $\ell = 2,\ m = 1$
  modes for all simulations as a function of time from merger ($t - t_0$). The
  data are multiplied by a factor of two to account for the symmetry between
  positive and negative $m$. A contribution from the $m=1$ mode is present in
  all simulations, especially at high resolution. \acp{GW} from the
  spiral mode are subdominant during most of the evolution, but show negligible
  damping.
  }
  \label{fig:E.GW.dot}
\end{figure*}

We estimate the strength of the one-armed spiral instability in an unambiguous
way from the multipole decomposition of the \ac{GW} energy and angular momentum
fluxes, $L_{\rm GW}$ and $\dot{J}_{\rm GW}$, at $\mathcal{J}^+$. These are
obtained from the spin $-2$ weighted spherical harmonics decomposition of the
strain $h_{\ell m}$ at future null-infinity following \cite{bernuzzi:12}.
The multipoles of the energy flux carried by \acp{GW} are
\begin{equation}
  \big(L_{\mathrm{GW}}\big)_{\ell m} =
  \big(\dot{E}_{\mathrm{GW}}\big)_{\ell m} =
  \frac{1}{16 \pi} |\dot{h}_{\ell m}|^2\,,
\end{equation}
while the multipoles of the angular momentum flux are
\begin{equation}
  \big(\dot{J}_{\rm GW}\big)_{\ell m} =
  \frac{m}{16\pi} \Im [ h_{\ell m} \dot{h}_{\ell m}^\ast ]\,.
\end{equation}
We show the $\ell=2,\ m=1$ and $\ell=2,\ m=2$ quadrupole modes of the \ac{GW}
energy flux in Figure~\ref{fig:E.GW.dot}. While the $\ell=2,\ m=2$ mode peaks at
merger and then decays over a timescale of several milliseconds, the $m=1$ mode
grows after merger and saturates within a few milliseconds for both the MS1b and
SLy binaries.  In both cases, the mode appears not to be damped by
hydrodynamical processes and persists for the entire duration of the
simulations, \ie, up to $50\ \mathrm{ms}$ after merger in the MS1b case, or
until \ac{BH} formation in the SLy case.

The energy released in \acp{GW} by the one-armed spiral instability is several
orders of magnitude smaller than that from the dominant $\ell=2,\ m=2$ mode
after the merger. The $m=1$ \ac{GW} emission is not dynamically
relevant for the evolution of the remnant in the first several tens of
milliseconds. Even on secular timescales (for the MS1b binary) the $m=1$ mode
does not appear to be efficient at removing angular momentum from the remnant
with $J/\big(2 \dot{J}_{\mathrm{GW}}\big)_{\ell=2,\, m=1} \gtrsim 100\
\mathrm{s}$. For this reason, the $\ell=2,\ m=1$ is only very weakly damped by
\ac{GW} backreaction.  This is in contrast with the behavior of the dominant
quadrupole mode, which is a highly efficient emitter of \acp{GW} and, for this
reason, it is strongly damped over a timescale of $\sim 10$~ms
\cite{bernuzzi:15c}. After the $\ell = 2,\ m=2$ mode has decayed, the $\ell=2,\
m=1$ mode becomes the most luminous mode. As shown in Figure~\ref{fig:E.GW.dot},
this happens already $20\ \mathrm{ms}$ after merger in the highest-resolution
MS1b simulation. The dominance of this mode over long timescales suggests that
the one-armed spiral instability might leave some imprint on the \ac{GW} signal
if it survives for sufficiently long time.

Figure~\ref{fig:E.GW.dot} also shows that numerical viscosity in low-resolution
simulations can prevent the one-armed spiral instability from fully developing.
One could speculate that one of the reasons why this instability has gone
undetected for long time is that it might have been suppressed in simulations
performed at lower resolutions and/or using more dissipative numerical schemes
than those used here. Another reason is the $\pi-$symmetry that was assumed in
many previous simulations, which obviously prevents the instability completely.
We also remark that, while the instability is very evident for stiff \ac{EOS}
(Figure~\ref{fig:rho.xy}), it is less so in the case of softer \ac{EOS} (e.g.,
in our SLy model). In the latter case, it is difficult to identify from the
inspection of density colormaps. A modal decomposition of the density
distribution, or an analysis of the \ac{GW} multipoles are necessary to
unambiguously reveal it.

\section{Hybrid Waveforms}
We construct the complete \ac{GW} signal in the Advanced LIGO band from our
binaries by hybridizing our \ac{NR} waveforms with the tidal \ac{EOB} model
presented in \cite{bernuzzi:15a}. We generate \ac{EOB} waveforms using a
publicly available code \cite{eobcode}, starting at a frequency of $\simeq 10\
\mathrm{Hz}$, corresponding to $\simeq 18$ minutes before merger, and extending
up to the moment of merger \cite{bernuzzi:14b, bernuzzi:15a}. We include
multipoles up to $\ell = 4$. The resulting hybridized waveforms are publicly
available~\cite{waveforms}.

We align \ac{NR} and \ac{EOB} data as in \cite{bernuzzi:15a}.  We use the
difference between the two highest resolutions as a conservative estimate for
the numerical uncertainty in the $\ell=2,\ m=2$ \ac{GW} phase predicted by our
simulations. The difference is less than $3$~radians at merger and less than
$0.3$~radians in the time window where we perform the alignment with \ac{EOB}.
Excluding the lowest resolution simulation, which appears not converged, we find
better than third order convergence in the phase and amplitude of the $\ell=2,\
m=2$ \ac{GW} mode for both the MS1b and the SLy binaries until shortly before
merger. This is similar to what was reported in \cite{radice:13a}. As a
consequence, the de-phasing between the \ac{NR} and \ac{EOB} waveforms is
dominated by the residual orbital eccentricity and the residuals are essentially
flat until shortly before merger as in \cite{bernuzzi:15a}.

In order to be able to estimate the detectability of the one-armed spiral mode,
we extend the $\ell=2,\ m=1$ \ac{GW} signal of the MS1b binary to $\simeq 1$
second after merger using a simple damped sinusoid. This is justified by the
fact that the $\ell=2,\ m=1$ \ac{GW} signal from our MS1b simulations has a very
stable instantaneous frequency and a narrow spectrum. Since the $\ell=2,\ m=1$
\ac{GW} amplitude in our simulations shows low frequency oscillations of unclear
origin, we are not able to reliably estimate damping times from our data.
Instead, we heuristically set the damping time of the $\ell=2,\ m=1$ mode in our
hybrid waveform to $100\ \mathrm{ms}$. This value is consistent with the
amplitude evolution of the $h=0.15\ M_\odot$ run, but somewhat smaller than what
could be inferred from the $h=0.1\ M_\odot$ data. However, other physical
processes, such as neutrino cooling and angular momentum redistribution due to
magneto-turbulence, will likely become dominant over such timescales
\cite{hotokezaka:13} and might damp the one-armed spiral instability
\cite{fu:11,franci:13,muhlberger:14}. As a consequence, our estimate of the
survival time of the $m=1$ mode should constitute a reasonable upper limit.

\section{Detectability}

\begin{figure*}[t]
  \includegraphics[width=\columnwidth]{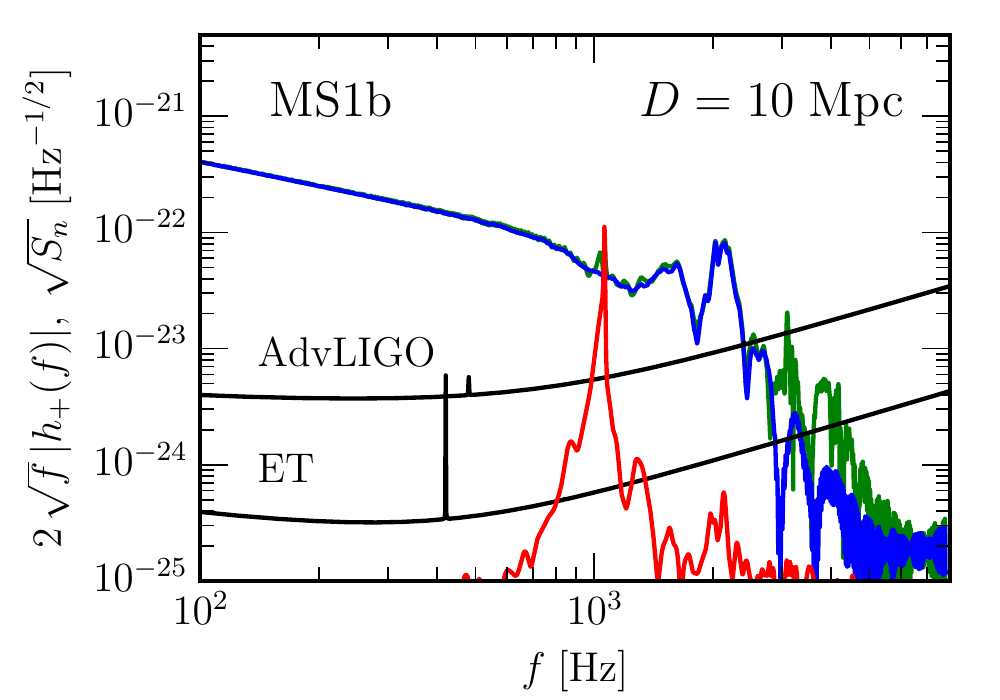}
  \includegraphics[width=\columnwidth]{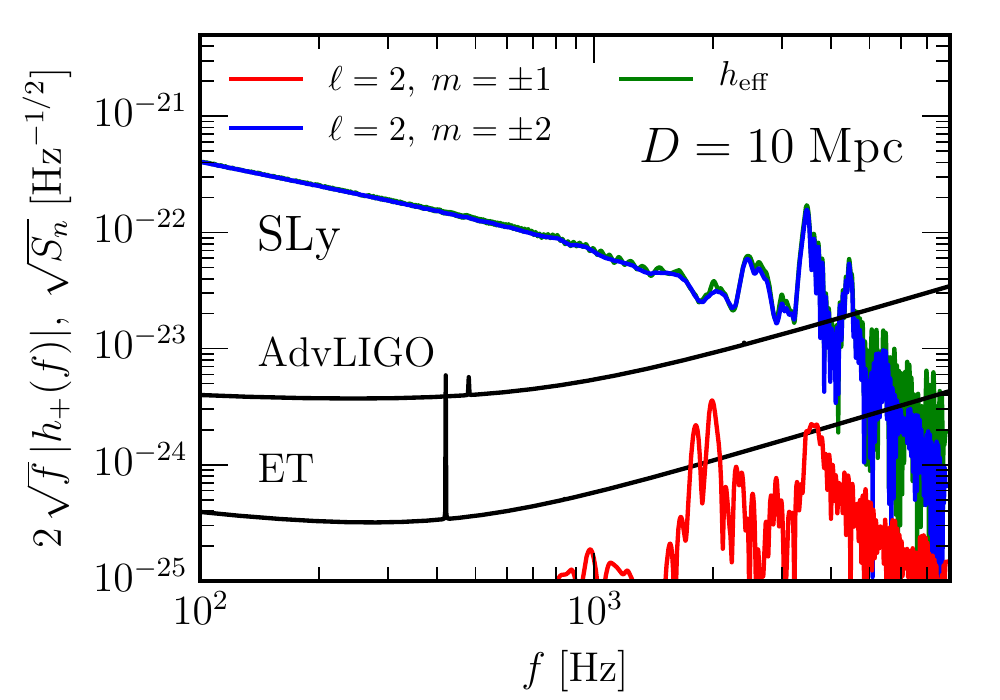}
  \caption{Spectrum of the effective \ac{GW} signal seen edge-on from a distance
  of $10\ \mathrm{Mpc}$. We show the spectrum of the hybrid waveforms
  constructed from the highest resolution data and the noise curves of Advanced
  LIGO and \ac{ET}. The $\ell=2,\ m=1$ spectrum is mostly concentrated in a
  narrow maximum located at half of the frequency of the $\ell=2,\ m=2$ peak.}
  \label{fig:SNR}
\end{figure*}

In Figure \ref{fig:SNR}, we show the spectrum of the effective \ac{GW} signals
for the highest resolution MS1b and SLy hybrid waveforms. For our analysis, we
assume an optimistic distance of $10\ \mathrm{Mpc}$ to the source, an optimal
sky location, and an edge-on orientation of the binary with respect to the
detector, which is optimal for the detection of the $\ell=2,\ m=1$ mode. We also
superimpose the sensitivity curves of Advanced LIGO in its zero-detuning
high-laser-power configuration \cite{LIGO-sens-2010} and of the proposed \ac{ET}
\cite{punturo:10, hild:11}.

We find the spectrum of the \ac{GW} signal generated by the one-armed spiral
instability to reach its maximum at a frequency half of that of the dominant
quadrupole peak $f_2$. It is thus conceivable that the detection of \acp{GW}
from the $m=1$ instability could be used to constrain the \ac{NS} \ac{EOS},
since $f_2$ has been shown to encode properties of the \ac{EOS} at high
densities \cite{bauswein:11, bauswein:14, bernuzzi:15b}.

We quantify the detectability of the different components of the \ac{GW} signal
by computing optimal \acp{SNR}, \ie, assuming an optimal detection template
\cite{sathyaprakash:09}, for Advanced LIGO and \ac{ET} using the hybrid
waveforms obtained from the highest resolution \ac{NR} simulations. We compute
the SNR integrals using $h_+(f)$ over the frequency windows $9\ \mathrm{Hz} \leq
f \leq 8192\ \mathrm{Hz}$ and $1\ \mathrm{Hz} \leq f \leq 8192\ \mathrm{Hz}$ for
advanced LIGO and \ac{ET} respectively. We estimate low-frequency ($f \lesssim
12\ \mathrm{Hz}$) contributions to the \ac{SNR} by extending the hybrid spectrum
as a power-law with index $-7/6$ at low frequencies \cite{maggiore:08}. The
results of this analysis are reported in Table \ref{tab:snr}.

\begin{table}
  \centering
  \caption{Single detector \acp{SNR} of different components of the \ac{GW}
  signal for the MS1b and SLy binaries at a distance of $10\ \mathrm{Mpc}$, seen
  edge-on, and computed assuming optimal sky location. We quote the \ac{SNR} of
  the entire waveform (all multipoles up to $\ell = 4$), the \ac{SNR} of the
  $\ell=2,\ m=1$ mode $\SNR_{2,1}$, and the \ac{SNR} accumulated by the
  $\ell=2,\ m=1$ mode in a window of $10\, \mathrm{ms}$ after merger
  $\SNR_{2,1}^{10\, \mathrm{ms}}$.  For the time windowing, we use $10\,
  \mathrm{ms} \leq t - t_0 \leq 20\, \mathrm{ms}$ for the MS1b binary and $3\,
  \mathrm{ms} \leq t - t_0 \leq 13\, \mathrm{ms}$ for the SLy binary. Finally,
  we also give the \ac{SNR} for the $\ell=2,\ m=2$ mode computed using only the
  high-frequency ($f \geq 1\ \mathrm{kHz}$) component of the \ac{GW} signal
  $\SNR_{2,2}^{f \geq 1\, \mathrm{kHz}}$.\vspace{1em}}
  \label{tab:snr}
  \begin{tabular}{ll|cccc}
  \hline
    Det. & Binary & $\SNR$ & $\SNR_{2,1}$ &
    $\SNR_{2,1}^{10\, \mathrm{ms}}$ &
    $\SNR_{2,2}^{f \geq 1\, \mathrm{kHz}}$ \\
  \hline
  LIGO    & MS1b-M135 & $\pz169.4$ & $\pz1.6$ & $0.62$ & $\pz5.4$ \\
  LIGO    & SLy-M135  & $\pz169.5$ & $\pz0.1$ & $0.09$ & $\pz6.9$ \\
  \hline
  ET      & MS1b-M135 & $2460.5$   & $14.4$   & $5.86$ & $47.4$ \\
  ET      & SLy-M135  & $2461.6$   & $\pz1.0$ & $0.80$ & $61.3$ \\
  \hline
 \end{tabular}
\end{table}

Despite its large spectral peak amplitude, the detection of the $m=1$ mode with
current laser interferometers appears unlikely. Even at the relatively close
distance of $D = 10\ \mathrm{Mpc}$, the optimal \ac{SNR} for the $\ell=2,\ m=1$
\ac{GW} mode of the hybrid MS1b waveform is only $\simeq 1.6$ for advanced LIGO.
For comparison, the threshold on the optimal \ac{SNR} for detection is typically
set to $8$ \cite{abadie:10}. This means that the one-armed spiral instability
will be undetectable even for nearby events and the $\ell=2,\ m=2$ \ac{GW} mode
appears much more promising for  detecting post-merger \acp{GW} from \ac{BNS}
mergers. This is even more so for the short lived merger remnant of the SLy
binary.

Table \ref{tab:snr} also reports the \ac{SNR} accumulated by the
$\ell=2,\ m=1$ \ac{GW} mode over $10\, \mathrm{ms}$ period after its amplitude
has saturated.  This value can be used to evaluate the dependency of the
\ac{SNR} on the survival time of the one-armed mode. Assuming that the $\ell=2,\
m=1$ \ac{GW} mode survives for a time $T$ with no damping, the total \ac{SNR}
for the MS1b binary, for example, can be computed as
\begin{equation}
  \mathrm{SNR} = 0.62\, \left(\frac{10\, \mathrm{Mpc}}{D}\right)\,
  \left(\frac{T}{10\, \mathrm{ms}}\right)^{1/2}\,.
\end{equation}
Our fiducial case, with a damping time scale of $100\, \mathrm{ms}$, would
correspond to an \emph{effective} survival time $T_{\rm eff} \simeq 67\,
\mathrm{ms}$.

We also remark that our analysis refers to the case of binaries seen edge-on.
In the face-on case the \ac{SNR} for the $\ell=2,\ m=2$ mode is twice as large,
while the $\ell=2,\ m=1$ mode is completely suppressed.

\section{Discussion}
In combination with previous studies by others \cite{bernuzzi:14a, kastaun:15,
paschalidis:15, east:16} our results for equal-mass, irrotational \ac{BNS}
mergers from quasi-circular orbits strongly suggest that the one-armed spiral
instability is a generic outcome of the merger of two \acp{NS}.  As we
demonstrate in the cases of both soft and stiff \ac{EOS}, even tiny asymmetries,
necessarily of numerical origin for exactly equal mass systems, but expected in
any astrophysical configuration, are sufficient to trigger the growth of this
instability. Once seeded at the time of merger, the one-armed spiral instability
quickly grows into a large-scale $m=1$ spiral density perturbation, which
saturates within a time scale of $\sim 10\ \mathrm{ms}$. Our results are
supported by the analysis of well defined, gauge invariant, quantities at
$\mathcal{J}^+$, and by a resolution study.

We find that \acp{GW} excited by the $m=1$ mode carry relatively little energy
and angular momentum as compared to those of the dominant $m=2$ mode. As a
consequence, the $m=1$ mode is very weakly damped and may persist over secular
timescales, while the $m=2$ mode decays over a timescale of $\sim 10$
milliseconds.

The characteristic frequency of the \acs{GW} emitted by the one-armed spiral
instability encodes important aspects of the \ac{NS} \ac{EOS}.  Unfortunately,
as our analysis shows, the direct observation of \acp{GW} from the $m=1$ mode by
the current \ac{GW} detectors appears unlikely.  We find that, even using an
optimal \ac{SNR} detection threshold as low as $5$, Advanced LIGO at its design
sensitivity will be able to detect optimally oriented sources only out to $\sim
3\ \mathrm{Mpc}$. The expected event rate for \ac{BNS} mergers in this volume is
only $10^{-7}\, \mathrm{yr}^{-1} - 10^{-4}\, \mathrm{yr}^{-1}$ \cite{abadie:10}.
This picture could only change if the $\ell=2,\ m=1$ mode is somehow able to
survive for many hundreds of milliseconds with no significant damping. However,
the decay rates we observe in our simulations seem to exclude this possibility.
In contrast, we find a horizon distances for the post-merger $m = 2$ mode of
$\sim 20\ \mathrm{Mpc}$ and $\sim 27\ \mathrm{Mpc}$ for the MS1b and SLy
binaries assuming optimal orientation for the $\ell=2,\ m=2$ mode.  This is in
agreement with the more careful analysis of results of conformally flat
simulations by \cite{clark:15}. \acp{GW} from the one-armed spiral instability
will be a target for third-generation detectors, such as \ac{ET}. For the
latter, we find an optimal \ac{SNR} for the $\ell=2,\ m=1$ mode of our highest
resolution MS1b hybrid waveform of $\simeq 14.4$ at $10\ \mathrm{Mpc}$ using the
ET-D sensitivity curve \cite{hild:11}. This would put the horizon for the
detection of an optimally oriented source at $\sim 29\ \mathrm{Mpc}$.  This
corresponds to an increase in the event rate by a factor $10^3$ with respect to
that of Advanced LIGO.

The most important limitation of our study is the omission of
magnetohydrodynamic effects.  Very strong magnetic fields have been shown to be
able to suppress this instability in isolated \acp{NS} \cite{franci:13}.
Muhlberger et~al.~\cite{muhlberger:14}, however, found that large scale
hydrodynamical instabilities in isolated differentially rotating \acp{NS} are
not affected and in some cases are even amplified by the presence of magnetic
fields over a large range of magnetic field strengths.

As a side product of the present study, we constructed, for the first time,
high-quality hybrid waveforms employing state-of-the-art analytical models and
high-resolution high-order \ac{NR} data. These waveforms are publicly
available~\cite{waveforms}.

\begin{acknowledgments}
We thank Stefan Hild for the ET-D noise curve data and acknowledge useful
discussions with L.~Baiotti, W.~E.~East, F.~Galeazzi, W.~Kastaun, K.~Kiuchi,
V.~Paschalidis, L.~Rezzolla, M.~Shibata, and K.~Takami.  This research was
partially supported by the Sherman Fairchild Foundation, by the International
Research Unit of Advanced Future Studies, Kyoto University, and by NSF under
award Nos.\ CAREER PHY-1151197, PHY-1404569, and AST-1333520. The simulations
were performed on the Caltech computer Zwicky (NSF PHY-0960291), on NSF XSEDE
(TG-PHY100033), and on NSF/NCSA Blue Waters (NSF PRAC ACI-1440083). This article
has been assigned Yukawa Institute report number YITP-16-21.
\end{acknowledgments}

\acrodef{BH}{black hole}
\acrodef{BNS}{binary neutron stars}
\acrodef{HMNS}{hypermassive neutron star}
\acrodef{EOB}{effective-one-body}
\acrodef{EOS}{equation of state}
\acrodef{ET}{Einstein Telescope}
\acrodef{GR}{general relativistic}
\acrodef{GW}{gravitational wave}
\acrodef{NR}{numerical relativity}
\acrodef{NS}{neutron star}
\acrodef{SGRB}{short gamma-ray bursts}
\acrodef{SNR}{signal-to-noise ratio}

\bibliographystyle{apsrev}
\bibliography{references}

\end{document}